\documentstyle[preprint,aps]{revtex}
\tightenlines
\newcommand{\be}{\begin{equation}}
\newcommand{\ee}{\end{equation}}
\newcommand{\ben}{\begin{eqnarray}}
\newcommand{\een}{\end{eqnarray}}
\begin{document}

\title{Chaotic behavior in a $Z_2\times Z_2$ field theory\thanks{This
work is supported in part by funds provided by the U.S. Department of
Energy (D.O.E.) under cooperative research agreement \#DF-FC02-94ER40818.}}

\author{V. Latora$^{1,2}$ and D. Bazeia$^{1\,}$\thanks{On leave from
Departamento de F\'\i sica, Universidade Federal da Para\'\i ba,
Caixa Postal 5008, 58051-970 Jo\~ao Pessoa, Para\'\i ba, Brazil}}

\address{$^1$Center for Theoretical Physics, Laboratory for Nuclear Science
and Department of Physics\\
Massachusetts Institute of Technology, Cambridge, Massachusetts 02139}

\address{$^2$Department of Physics, Harvard University, Cambridge,
Massachusetts 02138}

\maketitle

\begin{abstract}
We investigate the presence of chaos in a system of two real scalar fields
with discrete $Z_2\times Z_2$ symmetry. The potential that identify the
system is defined with a real parameter $r$ and presents distinct features
for $r>0$ and for $r<0$. For static field configurations, the system
ntsupports two topological sectors for $r>0$, and only one for $r<0$.
Under the assumption of spatially homogeneous fields,
the system exhibts chaotic behavior almost everywhere in parameter space.
In particular a more complex dynamics appears for $r>0$; in this case chaos
can decrease for increasing energy, a fact that is absent for $r<0$.
\end{abstract}

\newpage

\section{Introduction}

Nonlinearity plays an important role in field theory. It is
responsable for the presence of interactions and may enter the game
allowing interesting situations, as is the case for instance when the system
engenders spontaneous symmetry breaking. In this case the classical equations
of motion may give rise to interesting field configurations such as kinks,
vortices and monopoles, depending on the particular model in consideration.
See for instance Refs.~{\cite{raj82,rso84,vil94}} for details.

In the standard route to defect formation, one usually search for static
field configurations that solve the equations of motion and present finite
energy. In the simplest case of a single real scalar field, a system given
in terms of a potential that presents $Z_2$ symmetry may support topological
solutions in $1+1$ dimensions when the potential has at least two degenerate
minima. In the case of two real scalar fields, the system is richer and may
support distinct types of topological solutions, as we illustrate below.

Nonlinearity also plays an important role on chaotic behavior of systems.
For linear systems, the qualitative nature of the behavior does not change
when one changes their parameters. However, for systems governed by nonlinear
dynamics one may find examples where small change in a parameter can lead to
dramatic changes in both the quantitative and qualitative behavior of the
system. See for instance Refs.~{\cite{lli92,rei92,str94}} for details.

There are distinct routes to chaos in field theory, and here we shall follow
the point of view introduced in Ref.~{\cite{sav83}}. In this case one
investigates field theoretical models under the assumption of spatially
homogeneous field configurations. We think of field
configurations whose space variations are much smaller than the
corresponding time variations, and so we treat the fields as depending
only on time. This point of view has been explored in several different works,
as for instance in Refs.~{\cite{blm93,kaw95,sal95} and in some works therein.
We notice that these works deal with different models, describing Abelian
and non Abelian gauge fields in the presence
of spontaneous symmetry breaking. In these cases the symmetry
to be broken is continuum, local, and the system supports vortices or
monopoles. In the present work, however, we shall explore another system,
simpler, that presents discrete symmetry and is described by a couple of real
scalar fields. Recent investigations have shown that this system presents
interesting properties and some soliton solutions have also been found, as we
comment on below.

The main motivation of the present work is to investigate the presence of
chaos in a system that describes two real scalar fields engendering
the discrete $Z_2\times Z_2$ symmetry. We shall do this by comparing the case
where one supposes the field configurations to be spatially homogeneous
with the more familiar situation which considers static configurations.
The last case is appropriate for searching for soliton solutions, and we
comment on that in the next Sec.~{\ref{sec:general}}. We consider the
case of spatially homogeneous fields in Sec~{\ref{sec:chaos}} where we
search for chaotic behavior. As we also comment on, the subject of this paper
is related to recent issues \cite{ble98}, which ask for instance whether the
chaotic behavior found in flat spacetime persists during the cosmological
expansion. In this paper we deal with a model which is defined in flat 
spacetime and may be seen as the results
of an expanding FRW universe, valid under the approximation of very slow
expansion rate. Furthermore, the present investigation lands very naturally
to the context of hybrid inflation, where one requires real
scalars such as the inflaton field and at least another scalar field, which
couples to the inflaton field \cite{lin91} in a way similar to the one
we consider in this work. We end the paper by summarizing
the results in Sec.~{\ref{sec:comments}}.

\section{General considerations}
\label{sec:general}

The system we study in this paper is a $Z_2\times Z_2$ model of two 
real fields $\phi(x,t)$ and $\chi(x,t)$. The potential that
specifies the system is defined in terms of a single real parameter, $r$,
which controls the number of minimum energy states of that potential.
As we are going to show, when $r$ changes sign the potential changes form,
from the case with four minima $(r>0)$ to the case where only two minima
are present $(r<0)$. For $r>0$, spontaneous symmetry breaking appears
in both the $\phi$ and $\chi$ directions in the $(\phi,\chi)$ plane. For $r<0$
we have spontaneous symmetry breaking only in the $\phi$-field direction, and
this last situation is similar to the case of hybrid inflation where
a first-order nonthermal phase transition after preheating seems to be
present, as recently considered in Ref.~{\cite{kkl98}}.

The field theory that we consider is described by the Lagrangian density
\be
\label{l1}
{\cal L}=\frac{1}{2}{\partial_{\alpha}}\phi\,{\partial^{\alpha}}\phi+
\frac{1}{2}{\partial_{\alpha}}\chi\,{\partial^{\alpha}}\chi-V(\phi,\chi)
\ee
We are using standard notation, with $\hbar=c=1$, and $V(\phi,\chi)$ is the
potential, which one supposes to be given by
\be
\label{v1}
V(\phi,\chi)=\frac{1}{2}\left(\frac{\partial h}{\partial\phi}\right)^2+
\frac{1}{2}\left(\frac{\partial h}{\partial\chi}\right)^2
\ee
where $h=h(\phi,\chi)$ is a smooth function of the two fields $\phi$ and
$\chi$. Here it obeys
\be
\label{h1}
\frac{1}{\mu}\,h(\phi,\chi)={\tilde h}(\phi,\chi)=r\,\left(\frac{1}{3}\phi^3-
\phi\right)+\phi\,\chi^2
\ee
The parameter $\mu$ is real, with dimension of energy, and $r$ is
another real parameter, dimensionless.

In $1+1$ dimensions this system presents interesting soliton
solutions \cite{bds95}, which has been used in applications in condensed
matter \cite{brs96a} and in field theory \cite{brs96b,bba97,etb98}. The
equations of motion for $\phi=\phi(x,t)$ and $\chi=\chi(x,t)$ are given by
\ben
\phi_{tt}-\phi_{xx}+h_{\phi}\,h_{\phi\phi}+h_{\chi}\,h_{\chi\phi} &=& 0
\\
\chi_{tt}-\chi_{xx}+h_{\phi}\,h_{\phi\chi}+h_{\chi}\,h_{\chi\chi} &=& 0
\een
These are the equations we shall deal with in the following. For simplicity,
however, we rewrite them in terms of dimensionless variables
${\tilde{t}}=\mu\,t$ and ${\tilde x}=\mu\,x$ to get
\ben
\phi_{{\tilde t}{\tilde t}}-\phi_{{\tilde x}{\tilde x}}+{\tilde h}_{\phi}\,
{\tilde h}_{\phi\phi}+{\tilde h}_{\chi}\,{\tilde h}_{\chi\phi} &=& 0
\\
\chi_{{\tilde t}{\tilde t}}-\chi_{{\tilde x}{\tilde x}}+{\tilde h}_{\phi}\,
{\tilde h}_{\phi\chi}+{\tilde h}_{\chi}\,{\tilde h}_{\chi\chi} &=&0
\een

In the case of static fields we have $\phi=\phi({\tilde x})$ and
$\chi=\chi({\tilde x})$. We change ${\tilde x}\to y$, for simplicity, and now
the equations of motion become
\ben
\label{x1}
{\frac{d^2\phi}{dy^2}}&=&2\,r^2\,(\phi^2-1)\,\phi+2\,(r+2)\,\phi\,\chi^2
\\
\label{x2}
{\frac{d^2\chi}{dy^2}}&=&2\,(\chi^2-r)\,\chi+2\,(r+2)\,\phi^2\,\chi
\een
which are the equations we deal with when searching for soliton
solutions. It is interesting to see that these equations of motion (\ref{x1})
and (\ref{x2}) can be solved by configurations that obey the pair of
first-order differential equations
\ben
{\frac{d\phi}{dy}}&=&r\,(\phi^2-1)+\chi^2
\\
{\frac{d\chi}{dy}}&=&2\,\phi\,\chi
\een
Solutions that obey this pair of first-order equations are BPS solutions
\cite{bps75}. They are stable configurations that minimize the energy, as
explicitly shown in \cite{bds95,brs96a}. See Ref.~{\cite{etb98}} for further
comments on BPS solutions, and for showing explicitly that the system defined
by the potential (\ref{v1}) is the bosonic portion of a supersymmetric theory.

In the case of spatially homogeneous fields we have $\phi=\phi({\tilde t})$
and $\chi=\chi({\tilde t})$. We change ${\tilde t}\to t$, for simplicity, and
here we get the equations of motion
\ben
\label{t1}
{\frac{d^2\phi}{dt^2}}&=&2\,r^2\,(1-\phi^2)\,\phi-2\,(r+2)\,\phi\,\chi^2
\\
\label{t2}
{\frac{d^2\chi}{dt^2}}&=&2\,(r-\chi^2)\,\chi-2\,(r+2)\,\phi^2\,\chi
\een
These are the equations we have to deal with when searching for a chaotic
behavior in the time evolution. It is interesting to see that the above
equations have some analogies with the equations investigated in
Ref.~{\cite{kaw95}}. However, in Ref.~{\cite{kaw95} the system under
consideration is the Abelian-Higgs model. This is the relativistic
generalization of the Ginzburg-Landau theory of superconductivity, and it is
well known that it presents vortex solutions \cite{nol73}. The reason for the
similarity between eqs.~(\ref{t1}) and (\ref{t2}) and the equations
investigated in \cite{kaw95} seems to rely on the assumptions introduced
in Ref.~{\cite{kaw95}}.

We follow the aim of this paper, which is to investigate the chaotic behavior
of eqs.~(\ref{t1}) and (\ref{t2}), together with the study of the static 
solutions of eqs.~(\ref{x1}) and (\ref{x2}).  
We start presenting some general considerations regarding 
the static solutions. Our system is identified by the potential
$V(\phi,\chi)=\mu^2\,{\tilde V}(\phi,\chi)$, where
\be
\label{pot}
{\tilde V}(\phi,\chi)=\frac{1}{2}\,r^2\,(\phi^2-1)^2+
r\,(\phi^2-1)\,\chi^2+\frac{1}{2}\,\chi^4+2\,\phi^2\,\chi^2
\ee
The potential depends on the real parameter $r$, and then 
the behavior of the system is directly related to such 
parameter. The main characteristic of this potential, which certainly strongly 
affect both the static solutions and the chaotic time behavior, is its
different shape for $r$ positive and negative. 
In fact, when $r>0$ the potential contains four vacuum states in the 
$(\phi,\chi)$ plane, while for $r<0$ there are only two vacuum states, 
and in principle we expect a different chaotic behavior 
for the dynamical trajectories that follows in accordance with eqs.~(\ref{t1})
and (\ref{t2}), as well as distinct features for the static solutions. 
It is important to say that in the two particular cases when $r=1$ 
and when $r=-2$ the two fields decouples \cite{bba97} 
and so we expect no chaotic behavior in these two cases.
First we show that, from the point of view of the topological 
soliton solutions admitted by Eqs.~(\ref{x1}) and (\ref{x2}), 
the case $r>0,\,r\neq1$ is richer than the case of $r<0,\,r\neq-2$. 
When $r$ is positive, $r\neq1$, the four vacuum states are given by
\ben
v_1&=&(-1,\,0),\hspace{1cm}v_2=(1,\,0)
\\
v_3&=&(0,-\sqrt{r}),\hspace{1cm}v_4=(0,\sqrt{r})
\een
For $r$ negative, $r\ne-2$, we have the two vacuum states
\be
\bar{v}_1=(-1,\,0),\hspace{1cm}\bar{v}_2=(1,\,0)
\ee
As it was already shown \cite{bds95,brs96a}, the energy of
the stable BPS solutions are obtained according to the values of
${\tilde h}(\phi,\chi)$ in the vacuum states.

For $r$ positive, $r\ne1$ we name $h_i$ as the value of $h$ at the vacuum
state $v_i$, and so we have
\ben
h_1&=&(2/3)\,\mu\,r
\\
h_2&=&-(2/3)\,\mu\,r
\\
h_3&=&h_4=0
\een
This means that we have two type of BPS solutions, with energies
$E_i=|\mu|\,\varepsilon_i, i=1,2$, where
\ben
\varepsilon_1&=&\frac{4}{3}\,r,\hspace{1cm}(r>0,\,r\neq1)
\\
\varepsilon_2&=&\frac{2}{3}\,r,\hspace{1cm}(r>0,\,r\neq1)
\een
In the first case the solutions connect the vacuum states
$(\pm1,0)$ and in the second case pair of vacuum states belonging to
different axes. In the topological sector defined by the two vacuum
states $(-1,0)$ and $(1,0)$ there are pairs of analytical
solutions given by \cite{bds95}
\be
\label{pair1}
\phi(y)=-\tanh[r(y-{\bar y})],\hspace{1cm}\chi(y)=0
\ee
and by
\ben
\label{pair21}
\phi(y)&=&-\tanh[2(y-\bar{y})]
\\
\label{pair22}
\chi(y)&=&\pm\frac{\sqrt{r-2}}{\cosh[2(y-{\bar y})]}
\een
Here ${\bar y}$ stands for the center of the topological solution. The first
pair (\ref{pair1}) is valid for $r>0$ and the second one for $r>2$.
We see that the second pair of solutions (\ref{pair21}) and (\ref{pair22})
gets to the first pair (\ref{pair1}) in the limit $r\to2$.

For $r$ negative, $r\ne-2$, there is just one type of BPS
solution, with energy $\bar{E}_1=|\mu|\,\bar{\varepsilon}_1$, where
\be
\bar{\varepsilon}_1=\frac{4}{3}\,|r|,\hspace{1cm}(r<0,\,r\neq-2)
\ee
The first pair of solutions given by Eq.~(\ref{pair1}) is also a pair of
solutions in the case $r<0$.

It is clear that for $r>0,r\ne1$ the situation is richer than 
in the case $r<0,r\ne-2$, at least from the point of view of the 
above topological soliton solutions. In the first case the system 
admits two type of topological BPS solutions, while in the second case
it has only one. We also expect a difference between the two $r>0$ and
$r<0$ regimes to be present in the chaotic behavior of the
eqs.~(\ref{t1}) and (\ref{t2}).

Another interesting feature of this model concerns the masses
of the $\phi$ and $\chi$ fields. From the potential $V(\phi,\chi)$ we can
introduce the matrix
\be
\label{mat}
\left(\,
{ {V^v_{\phi\phi}}\hspace{.6cm}{V^v_{\phi\chi}} }
\atop
{ {V^v_{\chi\phi}}\hspace{.6cm}{V^v_{\chi\chi}} }
\,\right)
\ee
where $V_{\phi\phi}=\partial^2{V}/\partial\phi\partial\phi$ and so
forth. The superscript $v$ stands for substituting the fields for
their vacuum values once the derivatives are done. For the potential
introduced in Eq.~(\ref{pot}) we see that
\be
V_{\phi\chi}=V_{\chi\phi}=4\,{\mu}^2\,(r+2)\,\phi\,\chi
\ee
and so $V^v_{\phi\chi}=V^v_{\chi\phi}$ vanishes for every vacuum state,
despite the sign of $r$. This means that we can read the (square) mass
of each field directly from the matrix (\ref{mat}). We have at most two
different mass values. They are
\ben
m^2_{\phi}&=&4\,\mu^2\,r^2\hspace{1cm}m^2_{\chi}=
\,4\,\mu^2\hspace{1cm}(r>0, \,\,\phi\,\,{\rm axis})
\\
m^2_{\phi}&=&4\,\mu^2\,r\hspace{1cm}m^2_{\chi}=
4\,\mu^2\,r\hspace{1cm}(r>0, \,\,\chi\,\,{\rm axis})
\een 
and also
\be
\bar{m}^2_{\phi}=4\,\mu^2\,r^2\,\hspace{1cm}
\bar{m}^2_{\chi}=4\,\mu^2\hspace{1cm}(r<0, \,\,\phi\,\,{\rm axis})
\ee
where we inform the axis of the vacuum state used to obtain the respective
masses. We note that for $r<0$ the $\chi$ field develops no symmetry breaking,
and its mass does not depend on $r$. This is the way this specific
$Z_2\times Z_2$ system behaves, and this behavior may have interesting
connections with models of hybrid inflation
\cite{lin91,kkl98,cle96,stb95,gkl97}. We remark that there are
other $Z_2\times  Z_2$ systems, as for instance the ones investigated recently
in Ref.~{\cite{etb98}}, which are also defined with a single real parameter,
but that present distinct behavior, with different sets of masses and vacuum
states. They certainly lead to richer scenarios, and may give other
informations on the role the parameter $r$ plays in such models.

\section{Chaotic behavior}
\label{sec:chaos}

In this Section we investigate the chaotic behavior of the system
introduced in Sec.~{\ref{sec:general}}. We focus our
attention to the pair of Eqs.~(\ref{t1}) and (\ref{t2}). 
These equations can be seen as the equations of motion that
follow from the Hamiltonian
\be
H=\frac{1}{2}\,\dot{p}^2_{\phi}+\frac{1}{2}\,\dot{p}^2_{\chi}+
{\tilde V}(\phi,\chi)
\ee
They describe the motion of a classical particle in the 
bidimensional potential ${\tilde V}(\phi,\chi)$. We remark that this choice
of Hamiltonian makes the energy dimensionless. The equations of motion can be
written in first order form, in terms of the canonical variables
$(\phi,p_{\phi})$ and $(\chi,p_{\chi})$,
\ben
\label{eom1}
\dot{\phi}&=&\frac{\partial H}{\partial p_{\phi}}
=p_{\phi}
\\
\dot{p}_{\phi}&=&- \frac{\partial H}{\partial \phi}
=2\,r^2\,(1-\phi^2)\,\phi\,-2\,(r+2)\,\phi\,\chi^2
\\
\dot{\chi}&=& \frac{\partial H}{\partial p_{\chi}} 
=p_{\chi}
\\
\label{eom4}
\dot{p}_{\chi}&=&- \frac{\partial H}{\partial \chi}
=2\,(r-\chi^2)\,\chi-2\,(r+2)\,\phi^2\,\chi
\een 
In the rest of the paper we study for which values of the parameter $r$ 
and of the energy $E $ the trajectories
$\phi(t), \chi(t), p_{\phi}(t)$, and $p_{\chi}(t)$ show a regular or a
chaotic behavior. 

The motion is always bounded, i.e. confined to a finite region whose
size changes according to the energy $E $ of the particle. 
In Fig.~1 we report the potential contour plot defined by
${\tilde V}(\phi,\chi)=E$; the different values of the energy $E$ are
reported in the figure. Fig.~1a refers to $r=-5$ while Fig.~1b and Fig.~1c
refer to $r=1$ and $r=3$, respectively. 

For $r<0$ we note the presence of two minima 
$(\pm1,0)$, corresponding to the two vacuum states 
previously discussed, and a saddle point at the origin. 
The potential vanishes at the two minima, while at the origin $(0,0)$ it 
assumes the value $r^2/2$ (12.5 for case $r=-5$ in Fig.~1a).
Dynamical trajectories, e.g. solutions of Eqs.~(\ref{eom1})-(\ref{eom4})
are confined to one of the two minima regions if $E<12.5$, while they can
jump from one side to the other when $E  >12.5$. For $r=-2$ the two
fields decouple and the numerical investigation shows no chaotic behavior. 

The situation for $r>0$ is qualitatively different because we 
have four minima ${\bar V}=0$ at $(\pm1,0)$ and at $(0,\pm\sqrt r)$, 
corresponding to the four vacuum states previously discussed, 
and a local maximum ${\bar V}=r^2/2$ at the origin $(0,0)$. There are also
four saddle points with $\phi\neq0$ and $\chi\neq0$; they are given by
\be
\phi^2=\frac{r}{2\,(r+1)},\hspace{1cm}\chi^2=\frac{r^2}{2\,(r+1)}
\ee
In the saddle points the potential gets the value
\be
\label{sp}
{\bar V}=\frac{r^2}{2\,(r+1)}
\ee
The case $r=1$ is a particular case because the potential presents 
$Z_4$ symmetry \cite{bba97} (see Fig.~1b); the two fields 
decouple and the numerical simulations show no chaotic behavior.
When $r\ne1$ trajectories can have different
behaviors according to the energy $E$. We discuss the case $r=3$ in
Fig.~1c: when $E \le1.125$ the four minima are unconnected regions and 
the motion remains confined around the minimum in which we start the
trajectory. As soon as $E >1.125$ (see Fig.~1c) the trajectory
crosses the saddle point [see Eq.~(\ref{sp})] between the minima and wonders
from one minimum to another. The region around the relative maximum at the 
origin $(0,0)$ is still not allowed. 
When $E >r^2/2=4.5$ the dynamical trajectories can cross the 
origin and the four mimina are all directly interconnected.

In order to study the dynamical behavior of our system 
for different values of $r$ and energy $E$, we integrate 
numerically Eqs.~(\ref{eom1})-(\ref{eom4}) using a fourth order 
symplectic algorithm with a time step $\Delta t = 0.001$ \cite{yo}.
The time step has been determined in order to keep the error in 
energy conservation below ${\Delta E /E } =10^{-8}$ for 
any value of $r$ and $E $; the results are stable respect 
to a further reduction of $\Delta t$. For any conservative system such as
the one we are considering, 
the Hamiltonian is an integral of motion, and the energy conservation 
restricts trajectories to lie on a three-dimensional surface 
$H(\phi, \chi, p_{\phi}, p_{\chi})= E $ in the four-dimensional
phase space $(\phi, \chi, p_{\phi}, p_{\chi})$. 
We can obtain a graphical information about 
the system by plotting the intersection of this three-dimensional 
surface with a plane. These kind of plot is called Poincar\'e surface of
section, and gives an indication about the dynamical behavior of the
system. To define the surface of section in our case  
we follow a trajectory and we plot $\phi$ and
$\chi$ each time $p_{\phi}=0$. Regular regions will appear as a series of
points (a mapping) which lie on a one dimensional curve (invariant KAM curve), 
while chaotic regions will appear as a scatter of points limited 
to a finite area due to energy conservation \cite{lli92,rei92}.

In Fig.~2 and in Fig.~3 we show the surface of section in the plane
$(\phi,\chi)$ for $r=-5$ and for $r=3$, respectively.
Each figure is obtained from $100$ different trajectories 
followed for a time $t=500$. The trajectories are generated from random
uniformly-distributed initial conditions
$\phi_0,{p_{\phi}}_0,\chi_0,{p_{\chi}}_0$.

For $r=-5$ we plot the two cases $E=10$ and $E=20$ (Fig.~2a and 
Fig.~2b respectively). When $E =10$ the two minima are separated 
and the system exhibits only regular behavior 
(the surface of section shows only invariant KAM curves). 
When $E =20$ the two minima are connected and the surface 
of section contains both regular and stochastic regions. 
The volume of the stochastic region increases with increasing energy,
as it will be clear when we calculate Lyapunov exponents. 

For $r=3$ the situation is richer. We report the surface of 
section for three different energies $E=0.5$, $E=2$, and
$E=6$. When $E=0.5$ (Fig.~3a) the four mimina are not interconnected and 
we only have invariant KAM curves. When $E=2$ (Fig.~3b) neighbour 
minima regions are connected. Regular and stochastic regions cohexist 
in the surface of section and are interwingled in a
very complicated way. The scatter of points representing the
chaotic region fill the area between regular curves, and then smaller 
regular islands are imbedded in the chaotic sea. The white regions 
correspond to regular region, and these would be filled by regular curves 
if we could increase even more the number of initial conditions considered 
in the figure. For energy $E=6$ (Fig.~3c) the trajectory can cross the
center and the resulting surface of section shows an increase of the chaotic 
area respect to the case $E=2$.

The cases $r=1$ and $r=-2$ (which we do not report in figure) are two very 
particular cases. The two equations decouples, and indeed the Poincar\'e
surface of section show only KAM curves for every value of the energy $E$. 

The information we can get from Poincar\'e surface of section is
qualitative. A way to quantify the chaotic behavior of a system is by
calculating its Lyapunov exponents. Chaos is defined in terms of the dyamical
behavior of pairs of orbits which initially are close together in the phase 
space. The Lyapunov exponents are given by the rate of exponential divergence
of close orbits and are defined from the long term evolution of an initial
infinitesimal volume $\Omega$ in the phase space from the following formula
\be
\label{liap}
\lambda_i = \lim_{{t \rightarrow \infty}} \lambda_i(t) = 
            \lim_{{t \rightarrow \infty}} \frac{1}{t} 
            \ln \frac{d_i(t)}{d_i(0)}
\ee
where $d_i(t)$ are the lenght of the principal axes of the 
ellipsoid $\Omega$ at time $t$. 
There is one Lyapunov exponent for each dimension of the phase space 
and the flow is chaotic \cite{lli92,rei92}
if at least one Lyapunov exponent is positive. 
For a two-dimensional Hamiltonian system 
we expect two positive $\lambda_1, \lambda_2$ and two
negative $\lambda_3, \lambda_4$ Lyapunov exponents with the following 
symmetry $\lambda_3= -\lambda_2$ and $\lambda_4= -\lambda_1$, 
because of the symplectic structure of Hamiltonian systems. 

To compute Lyapunov exponents for different values of $r$ and energy
$E $ we use the standard method developped in Ref.~{\cite{ben}}. 
The average number of time steps to use in order
to have a good convergence of the results is about 
$10^7-10^8$, which means $t=10^4-10^5$.   
The symmetries $\lambda_3= -\lambda_2$ and $\lambda_4= -\lambda_1$ 
are always perfectly verified in the numerical results;  
this is a good check for the integration codes.    
Fig.~4 shows the numerical calculation of the two largest Lyapunov 
exponents for case $r=3$ and $E =5.5$; 
we report the quantity $\lambda_i(t)$ for $i=1,2$ as a function of time. 
The top panel in Fig.~4 shows two different trajectories both starting 
from initial conditions in the chaotic region. After a transient time,
$\lambda_1(t)$ converges for both trajectories to a
constant value of about $0.3$, while $\lambda_2(t)$ converges to $0.01$.
To obtain $\lambda_i$ we evaluate numerically the limit in formula
(\ref{liap}) considering a time average of $\lambda_i(t)$ over the time
interval $t=2 \cdot 10^4 - 3\cdot  10^4$. The bottom panel of Fig.~4 shows the
behavior of $\lambda_1(t)$ and $\lambda_2(t)$ for a different initial 
condition, this time chosen to be in the regular region of the phase space;  
$\lambda_1(t)$ and $\lambda_2(t)$ do not reach a finite asymptotic value but
tend to zero for large $t$, showing that the Lyapunov exponents are equal to
zero.

For each value of $r$ and energy we consider $N=100$ 
different trajectories with random initial conditions 
$\phi_0, {p_{\phi}}_0, \chi_0, {p_{\chi}}_0$. 
We compute $\lambda_i$ for each trajectory and we count  
the number $N_c$ of chaotic trajectories (i.e. with positive 
$\lambda_1$, as in Fig.~4a).  
In the following we report the ratio $R=N_c/N$ 
as an indicator of the measure of volume of phase 
space which is chaotic, and the value of $\lambda_1$ 
averaged over the $N_c$ chaotic trajectories. 
We do not report $\lambda_2$ which is, 
for each value of $r$ and energy, 
one order of magnitude smaller than $\lambda_1$.  

We consider separately the two cases $r<0$ and $r>0$ 
for which we expect to find a qualitative difference in the 
chaotic properties of Eqs.~(\ref{t1}) and (\ref{t2}). 
We discussed above the topological soliton solutions of 
Eqs.~(\ref{x1}) and (\ref{x2}) for these two cases 
and we found that case $r>0$ is richer than case $r<0$. 
In fact, for $r>0,r\ne1$ the system supports two topological sectors,
admitting two different type of topological BPS solutions; for
$r<0,r\ne-2$ there is just one topological sector. This difference is due to 
the presence of four or two minima in the potential $V(\phi,\chi)$,
for $r>0$ or $r<0$ respectively.
We expect the difference in the shape of the potential to have important 
implications also on the chaotic behavior of our system. 
In Fig.~5 and Fig.~6 we plot the ratio $R=N_c/N$ and $\lambda_1$, 
as function of energy for the two cases $r<0, r\ne-2$ and 
$r>0,r\ne1$, respectively. The system exhibits an order to chaos 
transition as a function of energy for any value of $r$, but for $r=-2$
and $r=1$. These two cases are particular cases
because the two field decouple and the system degenerates to two systems
of a single field each one; $R$ and $\lambda_1$ are identically equal to zero 
in the whole energy range and are not reported in figures.

We show the typical behavior for $r$ negative in Fig.~5. 
We report $R$ and $\lambda_1$ vs. energy for $r=-5$; other values of $r$ 
have the same qualitative behavior. The onset of chaos is for energy equal to  
$r^2/2$ ($12.5$ in figure), when both minima can be visited 
by a single trajectory and $R$ and $\lambda_1$ start to be different 
from zero. For larger values of energy both $R$ and $\lambda_1$ increase 
with the energy. 

The situation is different for $r$ positive.
In Fig.~6 we plot $R$ and $\lambda_1$ vs. $E $ for 
two different values of $r$, $r=1.5$ (dashed line) 
and $r=3$ (dashed-dotted line). 
The behavior for $r=1.5$ and $r=3$ is 
qualitatively similar, although shifted in energy. 
Chaos starts when the energy overcomes the barrier between two
minima [which is given in accordance with Eq.~(\ref{sp})], and it keep 
increasing as function of the energy until a new back bending 
appears, both in $R$ and in $\lambda_1$. 
This behavior is to be connected to a stabilization effect 
occurring when the trajectory can cross the origin since both 
$R$ and $\lambda_1$ have local minima at $E =r^2/2$.
Although the chaotic behavior for $r=1.5$ and for $r=3$ are qualitatively
similar, the absolute value of $R$ and $\lambda_1$ is as larger 
as closer $r$ gets to unity. This result can be of interest to application
since the absolute value of $r$ also controls the masses of the two fields
and may be used in models of hybrid inflation.

The behavior at $r>0$ is therefore different from the behavior at $r<0$.
The most important qualitative difference between these two
cases is that for $r>0$ we can find localized
regions in energy where suppression of chaotic behavior appears for
increasing energies. Such behavior is not present for $r$ negative, and
this is directly related to the reduction of the number of minima in this last
case. Similarly, we have shown that for static solutions the number of
topological sectors changes from two to one when $r$ changes from positive
to negative values, respectively.

\section{Comments and conclusions}
\label{sec:comments}

In this work we have investigated the presence of chaotic behavior in a system
of two real scalar fields that engenders the $Z_2\times Z_2$ symmetry.
The system is defined by a potential that is controlled by a single real
parameter, $r$. This parameter can be positive or negative, and each case
leads to different behavior. For static field configurations for instance,
the system suports two dintinct topological sectors for $r$ positive, $r\ne1$,
and only one when $r$ is negative, $r\ne-2$.

Within the context of chaotic behavior, we have shown that chaos is present
almost everywhere in parameter space, with distinct qualitative behavior
for $r>0$ and for $r<0$. The system is richer for $r$ positive, and this is
directly related to the presence of the four minima when $r>0$. 
For $r>0$ we can find localized regions in energy where suppression
of chaotic behavior appears for increasing energy values,
a fact that is absent for $r$ negative. On the other hand,
in the case of static solution there are two topological sectors
for $r>0$, in contraposition with the single sector that the
system supports for $r<0$.

There are other models that engender the $Z_2\times Z_2$ symmetry,
and that are also governed by a similar real parameter $r$. Some examples
appear in Ref.~{\cite{etb98}}, and they can also be
studied within the context of chaotic behavior, under the assumption
of spatially homogeneous field configurations. Such investigations can 
introduce further light toward a better understanding of the connection
between chaotic behaviors and the parameter $r$
that specifies the physical properties of the system.

As we have already commented on, the present investigations may be of
interest to inflationary cosmology, in connection with issues raised in
the recent works \cite{ble98,lin91,kkl98,cle96,stb95,gkl97}. The present
results seem to be valid in an expanding FRW universe when the rate of
expansion is very small. The chaotic behavior that we have
found may have been present in early times and may have played some role in
the cosmic evolution. A study in which the coupling to 
gravity is fully considered, and other related issues are presently 
under cosideration.

\acknowledgments
We would like to thank Michel Baranger for many interesting comments, 
and Giuseppe Politi for a critical reading of the manuscript. 
V.L. thanks BLANCEFLOR Foundation and CNR for financial support. 
We also thank Center for Theoretical Physics, MIT, and Department of 
Physics, Harvard University, for the kind hospitality.

\newpage
\centerline{{\bf FIGURE CAPTIONS}}

\bigskip
\noindent
Fig. 1.~
Potential contours plot for different values of E. 
a), b), and c) refer to $r=-5,1$ and $3$, respectively.

\bigskip
\noindent
Fig. 2.~
Poincar\'e surface of section for $r=-5$ in $(\phi,\chi)$ plane. 
$E=10$ in a) and $E=20$ in b).

\bigskip
\noindent
Fig. 3.~
Poincar\'e surface of section for $r=3$ in $(\phi,\chi)$ plane. 
$E=0.5$ in a), $E=2$ in b), and $E=6$ in c).

\bigskip
\noindent
Fig. 4.~
Lyapunov Exponents $\lambda_1(t)$ and $\lambda_2(t)$ 
vs. time for $r=3$ and $E=5.5$. Two chaotic trajectories are plotted in 
the top panel, while a regular trajectory 
is plotted in the bottom panel.

\bigskip
\noindent
Fig. 5.~
Fraction R of the phase space which is chaotic and largest 
Lyapunov exponent $\lambda_1$ vs. $E$ for $r=-5$  

\bigskip
\noindent
Fig. 6.~
Same as in Fig.5 for two positive values of $r$, namely $r=1.5$ (dashed)
and $r=3$ (dotted-dashed).

\end{document}